\documentclass[sigconf]{acmart}
\usepackage{diagbox}
\usepackage{multirow}
\usepackage{multicol}
\usepackage{listings}
\AtBeginDocument{%
  \providecommand\BibTeX{{%
    \normalfont B\kern-0.5em{\scshape i\kern-0.25em b}\kern-0.8em\TeX}}}

\copyrightyear{2023}
\acmYear{2023}
\setcopyright{acmlicensed}\acmConference[EASE '23]{Proceedings of the International Conference on Evaluation and Assessment in Software Engineering}{June 14--16, 2023}{Oulu, Finland}
\acmBooktitle{Proceedings of the International Conference on Evaluation and Assessment in Software Engineering (EASE '23), June 14--16, 2023, Oulu, Finland}
\acmPrice{15.00}
\acmDOI{10.1145/3593434.3593455}
\acmISBN{979-8-4007-0044-6/23/06}

\begin{document}

\title{HornFuzz: Fuzzing CHC solvers}

\author{Anzhela Sukhanova}
\email{bidelya@gmail.com}
\affiliation{%
  \institution{Saint-Petersburg University}
  \city{Saint Petersburg}
  \country{Russia}
}

\author{Valentyn Sobol}
\email{vo.sobol@mail.ru}
\affiliation{%
  \institution{Peter the Great St. Petersburg Polytechnic University}
  \city{Saint Petersburg}
  \country{Russia}
}

\begin{abstract}
Many advanced program analysis and verification methods are based on solving systems of Constrained Horn Clauses~(CHC). Testing CHC solvers is very important, as correctness of their work determines whether bugs in the analyzed programs are detected or missed. One of the well-established and efficient methods of automated software testing is fuzzing: analyzing the reactions of programs to random input data. Currently, there are no fuzzers for CHC solvers, and fuzzers for SMT solvers are not efficient in CHC solver testing, since they do not consider CHC specifics. In this paper, we present HornFuzz, a mutation-based gray-box fuzzing technique for detecting bugs in CHC solvers based on the idea of metamorphic testing. We evaluated our fuzzer on one of the highest performing CHC solvers, Spacer, and found a handful of bugs in Spacer. In particular, some discovered problems are so serious that they require fixes with significant changes to the solver.
\end{abstract}




\maketitle

\section{Introduction}
CHC solvers are widely used in static analysis. Constrained Horn Clauses~(CHC) are logical implications in first-order theories, and programs can be modeled as systems of such formulae~\cite{CHC}. There are high-performance CHC solvers that automatically solve these systems: Spacer~\cite{Spacer}, Eldarica~\cite{Eldarica}, PCSat~\cite{PCSat}, etc. Currently, the most well-known and efficient CHC solver is Spacer (which regularly performs well in the annual CHC solver competition, CHC-COMP~\cite{CHC-COMP-21}). This solver is part of the Z3~\cite{Z3} project and uses Z3 for SMT solving and interpolation.

If a CHC solver works incorrectly, it can lead to wrong conclusions during the analysis and, as a result, to undetected bugs. That is why it is necessary to look for bugs in CHC solvers. One of the fast and efficient approaches to finding bugs is fuzzing: an automated software testing technique that involves providing unexpected or random data as input to a computer program and analyzing the reaction of the program. It is commonly used in the domains of software security and quality assurance~\cite{FuzzSMT, BanditFuzz, STORM, Falcon, OpFuzz, YinYang}. 

To the best of our knowledge, there are no fuzzers for CHC solvers now, and fuzzers for SMT solvers~(SMT fuzzers) are not suitable for testing CHC solvers. Constrained Horn Clauses are formulae of a certain structure~\cite{CHC}, and SMT fuzzers usually do not retain this structure when they generate new inputs. In this scenario, we get not a CHC system, but an SMT formula, that is, we test not a CHC solver, but its SMT part. If SMT fuzzers retain the CHC structure, then they generate formulae with little to no variability, which is suboptimal for the fuzzing process. Moreover, the SMT fuzzers do not consider the peculiarities of CHC solver implementations.

Thus creating a fuzzer for testing CHC solvers, is of great interest. Some of the serious bugs that can occur in the solvers are incorrect satisfiability checks~\cite{STORM} and the generation of an invalid model. In this work, we have focused on finding exactly these bugs.

Our paper makes the following contributions:
\begin{enumerate}
    \item We propose to use metamorphic testing as a basis for fuzzing CHC solvers.
    \item We have designed and developed an open-source mutational fuzzer based on metamorphic testing for CHC solvers, HornFuzz\footnote{\url{https://github.com/AnzhelaSukhanova/HornFuzz} [accessed: \today]}.
    \item We have tested HornFuzz on Spacer CHC solver and successfully found both CHC system satisfiability bugs and some cases of wrong model generating. Some of the bugs have already been fixed by the Spacer developers. A group of problems with model generation has not yet been fixed, since it requires significant changes in the solver.
\end{enumerate}

The rest of the paper is organized as follows. In Sect. \eqref{sec:overview} we present the basic terminologies used throughout the paper and give an overview of our approach. In Sect. \eqref{sec:implementation} we explain the technical solutions and describe our implementation in detail. Then, in Sect. \eqref{sec:evaluation}, we analyze the bugs discovered. Sect. \eqref{sec:related} includes information about related work; we draw conclusions and briefly discuss plans for future work in Sect. \eqref{sec:conclusion}.

\section{Overview}\label{sec:overview}
This section gives a definition of Constrained Horn Clauses and discusses solvers of systems of such clauses. It also introduces fuzzing, metamorphic testing, and presents the idea on which HornFuzz is based.
\subsection{Constrained Horn Clauses}
A Constrained Horn Clause (CHC) is a first-order logic formula of the form
$\forall\,V (\varphi \wedge p_1[X_1] \wedge \dots \wedge p_n[X_n])\rightarrow h[X]$ , where
\begin{itemize}
	\item $\varphi$~--- constraint over some background theory;
	\item $V$~--- variables;
	\item $X_1,\,\dots X_n,\,X$~--- terms over $V$;
	\item $p_1,\dots,p_n$~--- uninterpreted fixed-arity predicates;
	\item $h$~--- uninterpreted fixed-arity predicate or $\bot$~\cite{CHC}.
\end{itemize}
The Constrained Horn Clause is linear if its premise contains at most one uninterpreted predicate. A system of clauses is linear if every clause in it is linear. Accordingly, a system is non-linear if at least one of its clauses is non-linear. Such systems of Constrained Horn Clauses are more difficult to solve than linear systems~\cite{Dima}. Rules are the Constrained Horn Clauses containing an uninterpreted predicate in the implication conclusion.

\subsection{CHC solvers}
To efficiently solve CHC systems, CHC solvers rely on performing multiple specific SMT queries, and to do that they use SMT solvers. SMT solvers are complex tools for evaluating the satisfiability of SMT instances. Satisfiability Modulo Theories (SMT) is the problem of deciding the satisfiability of a first-order formula over some first-order theories. 

In this paper, we will talk about the most widely used and efficient CHC solver~--- Spacer. Spacer is part of an open-source project Z3, one of the highest performing SMT solvers~\cite{Z3}, and uses its components for SMT-solving and interpolation~\cite{Spacer}. Spacer supports linear real and integer arithmetic, array theory, and offers best-effort support for many other SMT theories: data structures, bit-vectors, and non-linear arithmetic~\cite{CHC-tutorial}. It is also able to solve non-linear clauses.


\subsection{Fuzzing}
Fuzzing is a technique for automated software testing which is based on analyzing the program reaction to random input data~\cite{Fuzzing}. 

A fuzzer can be generation-based or mutation-based, depending on whether inputs are generated from scratch or by modifying existing inputs~\cite{Mutation-based}. In the first case, the fuzzer can generate data, for example, according to a specified grammar. Mutational fuzzers start their work with a certain set of initial inputs (so-called seeds). As they work, they change the seeds through the use of mutations.

Metamorphic testing is a variant of mutation-based fuzzing. It is a fuzzing technique that proposes to generate new test data while preserving some seed property~\cite{Metamorphic}. The expectation is that the seed and its mutant have a specific common property called the metamorphic relation; i.e., fuzzer mutations must retain this property.

\section{Concepts}
In this section, we present the main ideas on which HornFuzz is based. We describe the bug space that it considers and the mutations it uses.

\subsection{Main idea}
It is difficult to create a variety of Constrained Horn Clauses with non-trivial solutions, so the mutational approach is more suitable for fuzzing CHC solvers than generative. This way of generating test data is the base of HornFuzz. Additionally, there are many various benchmarks with CHC systems from solver competitions and papers that can be used as seeds.

At the moment, the capabilities of existing CHC solvers vary greatly, and having no reference solver is very beneficial. This discourages the use of solvers as oracles. Given this, metamorphic testing is of particular interest to find CHC satisfiability bugs. This testing technique does not require a complex oracle (in our case, another reference CHC solver) with which we would check the solver-under-test results.

In the context of satisfiability check bugs, the metamorphic relation is satisfiability. If the seed is satisfiable, then its metamorphic mutants should be satisfiable too and vice versa. Thus, using equivalent mutations, we can quickly check such a complex property as satisfiability. This idea formed the basis for our HornFuzz fuzzer.

\subsection{Bug space}\label{sec:Bug-space}
HornFuzz targets two kinds of bugs. First, it tries to find a satisfiability check bug. Each mutant is checked by the solver, and the result is compared with the seed satisfiability. If the satisfiability differs, we have found a satisfiability check bug.

If both formulae are satisfiable, the fuzzer substitutes the mutant's model (its satisfying assignment) into its formula and checks whether it is correct: every mutant clause should be satisfied by its model. If this property is violated, we have found a model generation bug.


Additionally, HornFuzz collects statistics on cases where CHC solver cannot solve the instance or timeouts when checking satisfiability or model. We describe the decision process for different cases in table~\eqref{tab:problem}.
\begin{table}[tb]
  \caption{Problem handling.}
  \label{tab:problem}
  \begin{tabular}{c|c|c|c} 
    \toprule
    \diagbox[width=9.7em]{Truth}{Solver result} & sat & unsat & unknown \\\midrule
    sat & check model & handle bug & \multirow{2}{*}{log info} \\\cmidrule{1-3}
    unsat & handle bug & pass &  \\\bottomrule
  \end{tabular}
\end{table}

\subsection{Mutations}\label{sec:mutations}
Mutations used by HornFuzz can be divided into three types: Z3 rewrites (mutation 9), changing solver parameters (mutation 10) and our own CHC specific mutations (mutations 1-8). Z3 rewrites and solver parameter transformations are complex satisfiability-preserving transformations which are already implemented in Z3. Mutations 1-4 and 6 can be applied only to the clause body.
\begin{enumerate}
	\item SWAP\_AND swaps two terms of the random conjunction:
	\begin{equation*}
		\varphi\wedge\psi\rightsquigarrow\psi\wedge\varphi
	\end{equation*}
	
	\item DUP\_AND duplicates one term of the random conjunction:
	\begin{equation*}
		\varphi\wedge\psi\rightsquigarrow\varphi\wedge\psi\wedge\varphi
	\end{equation*}
	
	\item BREAK\_AND splits the random conjunction into two:
	\begin{equation*}
		\varphi\wedge\psi\wedge\tau\rightsquigarrow\varphi\wedge(\psi\wedge\tau)
	\end{equation*}
	
	\item SWAP\_OR swaps two terms of the random disjunction:
	\begin{equation*}
		\varphi\vee\psi\rightsquigarrow\psi\vee\varphi
	\end{equation*}
	
	\item MIX\_BOUND\_VARS shuffles variables in the quantifier prefix:
	\begin{equation*}
		\forall\:(x, y, z).\psi(x, y, z)\rightsquigarrow\forall\:(y, z, x).\psi(x, y, z)
	\end{equation*}
	
	\item ADD\_INEQ replaces the random inequality with a conjunction of the same inequality and a less strong one:
	\begin{equation*}
		x<c\rightsquigarrow(x<c)\wedge(x<c+1)
	\end{equation*}
	
	\item ADD\_LIN\_RULE adds a linear rule that can be simplified to \begin{equation*}
	\forall\: \bar{x}.\bot \rightarrow P(\bar{x})
	\end{equation*}
	where $P$ is a randomly chosen uninterpreted predicate from the initial formula. The premise of the implication is one of the unsatisfiable formula set.
	
	\item ADD\_NONLIN\_RULE adds a non-linear rule of the form:
	\begin{gather*}
	\forall\:\bar{v}.(\exists\:\bar{x}.\ (x_1 > x_2\wedge P(\bar{x}, \bar{v}))\wedge (x_2 > x_3 \wedge\\
	P(\bar{x}, \bar{v}))\wedge\dots
	\wedge (x_n > x_1\wedge P(\bar{x}, \bar{v})))\rightarrow P(\bar {x}, \bar{v})
	\end{gather*}
	where $\bar{x} = (x_1, x_2,\dots, x_n)$, $x_i \in \mathbb{Z}$, $n$ is a random number from 1 to 10; $\bar{v} = (v_1, v_2,\dots, v_m)$, $m$~--- arity of a randomly chosen uninterpreted predicate $P$, all types of $\bar{v}$ elements correspond to the $P$ argument types.
 
    The notation $P(\bar{x}, \bar{v})$ actually means that $P$ is applied to a random sequence of $m$ arguments from the union of $\bar{x}$ and $\bar{v}$ with respect to the declared argument types (at least one such sequence always exists: it is $\bar{v}$).
	
	\item Equivalent rewrites offered by Z3 with or without parameters (see \eqref{app:simplify} for a complete list of parameters). 
	\item Changing the solver parameters (see \eqref{app:parameters} for a complete list of parameters) that affect the instance solving process~\cite{CHC}.
\end{enumerate}

CHC can be represented as $\forall\,V \neg(\varphi \wedge p_1[X_1] \wedge \dots \wedge p_n[X_n]) \vee h[X]$ and some seeds contain them in this form. We do not use mutations DUP\_OR and BREAK\_OR similar to mutations DUP\_AND and BREAK\_AND because doubling any of this disjunction terms would break the CHC structure.

Mutation MIX\_BOUND\_VARS can affect the order in which clauses are considered by the solver.

We expect the mutation ADD\_INEQ to affect the Model Based Projection~(MBP) process for LIA~\cite{MBP}. When MBP tries to eliminate integer variables from a formula, it builds upper and lower bounds for these variables. This bound is based on the formula inequalities and the choice of lower or upper bound depends on the number of the corresponding inequalities.

\section{Implementation}\label{sec:implementation}
In this section we demonstrate the details of HornFuzz implementation: give a description of HornFuzz workflow, talk about heuristics of seed selection and implementation details. This section also gives an overview of bug case reducer.

\subsection{HornFuzz process}
\begin{figure}[tb]
  \centering
  \includegraphics[width=\linewidth]{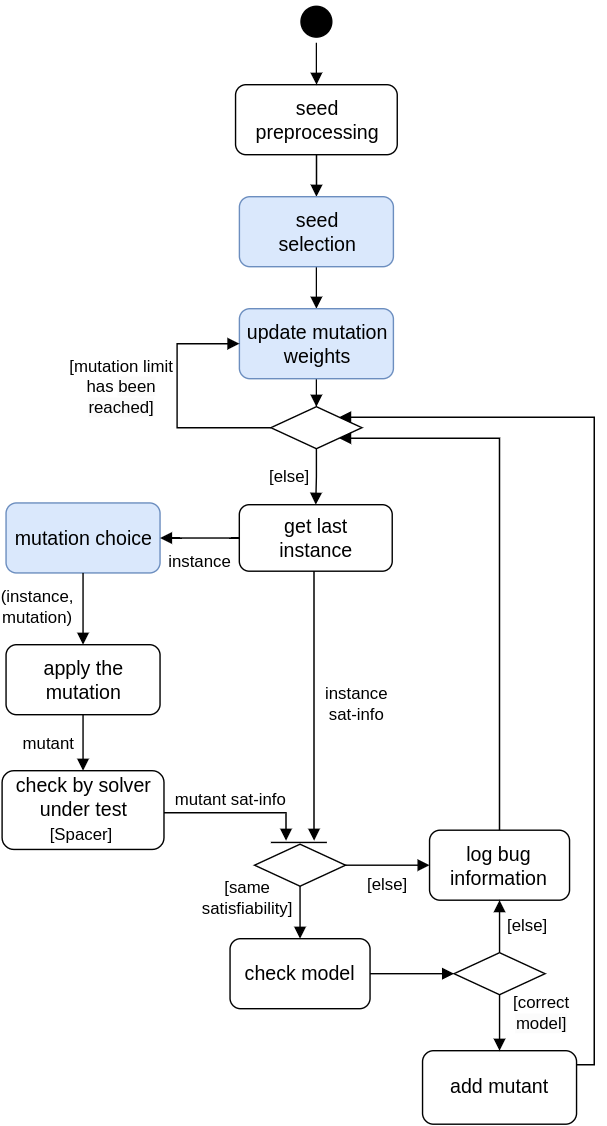}
  \caption{HornFuzz workflow.}\label{fig:arch}
  \Description{Activity diagramm describing the fuzzer work.}
\end{figure}

Figure~\eqref{fig:arch} presents an overview of the HornFuzz work process.

HornFuzz works continuously, mutating and checking instances until it is stopped forcibly. Its work begins with seed preparation. Each seed defines a group consisting of the seed and its mutated versions~(seed group). After seed preprocessing, a seed group is selected. The last group element will be mutated: it can be either the seed or its last mutant.

The next step is to select a mutation and apply it. Then the resulting mutant is checked for all types of interesting bugs handled (Section \eqref{sec:Bug-space}).

If no bugs are found, the mutant is added to its seed group. This process is then repeated until it becomes necessary to change the seed group. This can happen in the following cases.
\begin{itemize}
    \item The execution trace generated by the solver under test when solving mutants from the same seed group does not change for $5\cdot n$ where $n$~--- number of clauses in the system.
    \item Once a bug has been detected, continuing to work with that mutant's seed group may result in the same bug being rediscovered. Therefore, we use a bug detection limit upon reaching which the fuzzer proceeds to the next instance.
    \item We have reached a limit for the number of "unknown" solver results. This is done because there is a high probability that the solver will not be able to solve a mutant for a seed, which resulted in "unknown". For example, if an instance has become so complex that a solving timeout is reached, then most likely other mutations will also lead to a timeout.
\end{itemize}

Also in cases where the formula takes too long to be solved or the solver does not discover new traces for a long time, the mutants can be discarded. The seed group returns to the state before the addition of the mutants, that is, when it contains only the seed.

Highlighted blocks are configurable. They may be skipped or realized in several different ways, depending on the options with which HornFuzz is launched.
\begin{itemize}
    \item The choice of mutation can be weighted or equiprobable.
    \item Seed selection depends on the chosen heuristics, which attempt to predict which CHC instances are more likely to trigger a bug.
\end{itemize}

\subsection{Metrics}
There are two main metrics for fuzzer performance: code coverage~\cite{Coverage} and unique execution traces~\cite{Traces}. We use both metrics for different needs. 

To evaluate HornFuzz in the fuzzing process, we use the number of unique execution traces discovered when solving instances. This metric is good for tracking progress, since it allows us to understand whether the instance solving process has changed. Code coverage does not provide such information. 

In addition, it makes sense to collect not the entire execution trace, but the sequence of the main steps of solving the Constrained Horn Clause system. When a trace is detailed, small changes that do not affect the solution process (for example, deletion of subexpressions or other auxiliary actions) can lead to an increase in the number of unique traces. That is, with an insignificant change in the CHC system and its solution, the value of the metric will increase. To avoid such irrelevant boosting of this metric, we use the selective collection of unique execution traces. An example of the execution trace part is presented in the listing~\eqref{list:trace}.

For evaluating the effectiveness of the fuzzer in general, code coverage is a good metric. That is why we collect coverage statistics too. Unique execution traces are a priority metric for us, but it is also important to consider coverage statistics.

\begin{figure}[t!]
\begin{lstlisting}[
basicstyle=\small,
breaklines=true, 
caption=The execution trace fragment.,
captionpos=b,
frame=tb,
label=list:trace,
language=tcl,
numbers=left,
numbersep=4pt,
numberstyle=\tiny\color{darkgray}]
initialize muz/spacer/spacer_context.cpp:1585
init_rules muz/spacer/spacer_context.cpp:2420
add_rf muz/spacer/spacer_context.cpp:1036
add_rf muz/spacer/spacer_context.cpp:1068
assert_expr muz/spacer/spacer_prop_solver.cpp:126
assert_expr muz/spacer/spacer_prop_solver.cpp:126
assert_expr muz/spacer/spacer_prop_solver.cpp:126
reset muz/spacer/spacer_context.cpp:2365
log_expand_pob muz/spacer/spacer_context.cpp:3079
is_reachable muz/spacer/spacer_context.cpp:1340
internal_check_assumptions muz/spacer/spacer_prop_solver.cpp:335
is_reachable muz/spacer/spacer_context.cpp:1419
expand_pob muz/spacer/spacer_context.cpp:3576
expand_pob muz/spacer/spacer_context.cpp:3599
add_lemma muz/spacer/spacer_context.cpp:1966
log_add_lemma muz/spacer/spacer_context.cpp:3112
\end{lstlisting}
\end{figure}

\subsection{Seed selection}
Most mutation-based fuzzers are sensitive to seed selection~\cite{Selection}. Therefore, it is more efficient to choose an instance for mutation not in random order but according to some heuristic. HornFuzz implements several ways to prioritize CHC systems. When starting the fuzzer, any of the following three heuristics can be selected, as well as their combination (the instances can be divided into groups according to one heuristic and ordered in each group according to the other).

\begin{enumerate}
\item Selection of instances that cover the most rare transitions when solving.
\item Selection of the most complex instances. According to this heuristic, non-linear systems are preferred to linear ones, and within these groups they are ordered by the number of uninterpreted predicates.
\item Selection of the most simple instances, an inverse of heuristic (2).
\end{enumerate}

The second heuristic is designed to test the solver behavior on complex inputs, and the third heuristic is focused on increasing the number of launches because they will be faster.

Now we take a closer look at the first point, since this heuristic seems to us to be the most efficient in finding bugs. The solver can be represented as a system that changes its state at discrete times or discrete-time Markov chain~\cite{Markov_chain}. The states in this context are the steps of the CHC system solving recorded in the execution trace, and the transitions are given by two states following each other in the trace. That is, speaking of rare solver transitions, we mean pairs of states, not single state. It is important to understand how the solver got into a particular trace state, since this may indicate with what state of the solver or memory it came there. Therefore, information about the transition is more meaningful than information about visiting the trace state or line.

Moreover, this heuristic is particularly interesting because the transitions that the solver takes the least often correspond to parts of the solver that are executed rarely. Intuitively, rarely executed parts have a higher probability of containing a bug. Thus, this heuristic is not focused so much on opening new traces, but on reaching rare transitions, which are more likely to lead to the finding bugs.

We find the rarest transitions by collecting the statistics of solving the input formulae. HornFuzz builds a transition matrix for each instance and a combined matrix of all transitions using the execution traces. Thus, the probability of each transition can be calculated as the number of times this transition was made divided by the total number of transitions from its source state. 

Let each solver state have a number, $n$ be the total number of states. To select seed groups with the most rare transitions, we compile a transition matrix $P$. It is a stochastic matrix of transition probabilities $p_{ij}$ from state $i$ to state $j$. 
\begin{equation*}
P = \left(
\begin{array}{cccc}
p_{11} & p_{12} & \ldots & p_{1n}\\
p_{21} & p_{22} & \ldots & p_{2n}\\
\vdots & \vdots & \ddots & \vdots\\
p_{n1} & p_{n2} & \ldots & p_{nn}
\end{array}
\right)
\end{equation*}

From this matrix, a weight matrix $W$ can be obtained, where the transition weight is inversely proportional to the transition probability.
\begin{equation*}
w_{ij} =
  \begin{cases}
    \frac{1}{p_{ij}}\, , \; \text{if } p_{ij} > 0\\
    0 \quad \text{else}
  \end{cases}
\end{equation*}

By element-wise multiplying the weight matrix with the matrix $T_m$, which contains the number of transitions of a particular instance $m$, and taking the sum of all elements of the result, we get a value that indicates the priority of this instance:
\begin{equation*}
    k_m = \sum_{1 \leqslant i, j \leqslant n} w_{ij}\cdot t^m_{ij}
\end{equation*}
where $t^{m}_{ij}$ is an element of $T_m$. Instance selection is determined by the value of $k_m$.

\subsection{Mutation choice}
Depending on the HornFuzz launch options, it can choose mutations equiprobably or weighted. In both cases, the fuzzer first selects the type of mutation: rewrites, solver parameters, or own mutations. The probability of choosing one type or another is the same. The current mutation will be selected from mutations of this type.

If HornFuzz is run with the option of weighted mutation selection, the mutation weights are updated throughout the duration of the fuzzer work. The change in mutation weight depends on whether the mutation opens a new execution trace. Initially, all mutations have the same weight 0.1. 

The updated weight is calculated using the following formula based on the golden ratio.
\begin{equation*}
	w = 0.62\cdot w' + 0.38\cdot p
\end{equation*}
where $w'$~--- current weight and $p$~--- probability of opening a new trace (the ratio of how often this mutation resulted in a new unique execution trace to the total number of applications of this mutation).

\subsection{Reducer}
Large chains of mutations and complex instances are not suitable for reporting bugs as they are difficult for a human to understand. We implemented a custom test case reducer for our CHC fuzzing. It allows one to perform the following activities:
\begin{itemize}
    \item search for the minimum subsequence of the mutation chain, which still triggers a bug;
    \item simplify CHC system that triggered a bug~(problem CHC system).
\end{itemize}

Minimizing chains of mutations helps to localize the bug. In particular, if one of the mutations changes the solver parameter associated with a certain transformation, then most likely the bug is in this transformation. Reducing the mutatant greatly simplifies the bug analysis. For example, a formula that combines several theories after reduction may contain operations of only one theory, which allows you to quickly localize the bug.

For the mutation chain reduction the Delta Debugging algorithm is used~\cite{Reduction}, which can be described as follows. The mutation chain is divided into chunks of fixed size (initialized to half the size of the initial chain). Algorithm considers each chunk in turn, checking whether the bug still occurs when that chunk is removed, and eliminating it if so. Afterwards, the chunk size is halved and these steps are repeated. Reduction terminates when no chunk of size one can be removed.

The problem CHC system is reduced by removing subexpressions using the Hierarchical Delta Debugging algorithm~\cite{HDD}. Instead of reducing subsequences like regular Delta Debugging, it removes subtrees of the CHC system AST. First, an attempts are made to completely remove each clause from the system. Then the reducer goes through the AST of the remaining clauses. When excluding parts of an instance AST, the following must be true:
\begin{itemize}
    \item reduced system remains equivalent to its original formula;
    \item the bug remains reproducible.
\end{itemize}
Equivalence is checked as follows. Let $\{f_i\}_{i=0}^n$ be the original system of clauses and $\{r_i\}_{i=0}^n$ be the simplified one. Then the system $\{f_i\neq r_i\}_{i=0}^n$ must be unsatisfiable.

\subsection{Details}
We implemented the HornFuzz algorithm as a prototype fuzzer, written in Python. Below are some empirically derived constants used by the prototype.
\begin{itemize}
    \item The number of times the fuzzer works with instances from one seed group in a row is limited to 100.
    \item The mutation weights are recalculated every 1000 runs. 
\end{itemize}

The fuzzer can be launched with the following options.
\begin{itemize}
    \item \texttt{-mutations} allows one to choose mutation types, that is, HornFuzz can be configured to use only one or two mutation types (by default, HornFuzz uses all types). There are three mutation types: our own mutations, Z3 rewrites and changing solver parameters.
    \item \texttt{-heuristic} allows one to choose the seed selection heuristic. One of the following four heuristics can be chosen: seed selection by complex or simple inputs, by rare transitions, and in default order.
    \item \texttt{-options} allows to run the fuzzer with an equiprobable choice of mutations.
\end{itemize}

\section{Evaluation}\label{sec:evaluation}

\begin{table*}[tb]
\begin{center}
  \caption{Coverage and unique trace statistics averaged over all runs along with standard deviations.}
  \label{tab:coverage}
  \begin{tabular}{l|c|c|c|c|c}
  \toprule
    Data & Only seeds & Naive version & Transition version & Parameter version & Full version \\
   \midrule
    Runs & 3404 & \textbf{146083}~$\pm$~1781 & 120823~$\pm$~2828 & 52885~$\pm$~3942 & 47304~$\pm$~1968\\
    \midrule
    Line coverage, lines & 52257 & 52845~$\pm$~358 & 53380~$\pm$~553 & 56280~$\pm$~3100 & \textbf{57530}~$\pm$~566\\
    \midrule
    Line coverage, \% & 18.5 & 18.71~$\pm$~0.0013 & 18.9~$\pm$~0.002 & 19.92~$\pm$~0.011 & \textbf{20.37}~$\pm$~0.002 \\
    \midrule
    Growth, lines & $-$ & 588~$\pm$~26 & 1123~$\pm$~458 & 4023~$\pm$~2566 & \textbf{5273}~$\pm$~486\\
    \midrule
    Growth, \% & $-$ & 0.21~$\pm$~0.01 & 0.4~$\pm$~0.16 & 1.42~$\pm$~0.91 & \textbf{1.87}~$\pm$~0.17\\
    \midrule
    Unique traces & 933 & \textbf{12273}~$\pm$~138 & 12249~$\pm$~89 & 8991~$\pm$~1035 & 8461~$\pm$~265\\
    \midrule
    Number of runs to & \multirow{2}{*}{$-$} & \multirow{2}{*}{11.9} & \multirow{2}{*}{9.86} & \multirow{2}{*}{5.88} & \multirow{2}{*}{\textbf{5.59}}\\
    open a new trace &&&&&\\
    \bottomrule
  \end{tabular}
\end{center}
\end{table*}	

In this section, we talk about the HornFuzz evaluation and analyze its results together with the bugs found in Spacer CHC solver. We noticed that after adding mutations that use the solver parameters, the fuzzer began to detect more bugs. In addition, it is necessary to check the assumption about the efficiency of seed selection by rare transitions. Thus, we consider two main hypotheses: using solver parameters significantly increases the probability of finding a bug; using seed selection by rare transitions increases the fuzzer efficiency.

To test these hypotheses we aimed to answer the following research questions.

\textbf{RQ 1:} Does using solver parameters or/and seed selection by rare transitions allow one to explore different scenarios of solving the CHC systems?

\textbf{RQ 2:} Which mutations are most effective?

\textbf{RQ 3:} How effective are different HornFuzz configurations in finding bugs?

We run all experiments on a machine with the following environment: Arch Linux x86\_64 operating system, Intel Core i7-4790 CPU, 3.60GHz, 32Gb RAM.
\subsection{Seeds}
Our experiment uses the following CHC systems as seeds:
\begin{itemize}
    \item benchmarks of the CHC solver competition CHC-COMP for 2021~\cite{CHC-COMP-github};
    \item benchmarks of the international software verification competition SV-COMP~\cite{SV-COMP-gitlab};
    \item benchmarks from papers~\cite{Benchmarks1, Benchmarks2}.
\end{itemize}
HornFuzz does not use formulae that Spacer cannot solve or that cause a timeout. Currently, the fuzzer uses 3404 Constrained Horn Clause systems in LIA, LRA, array theory, and combinations.

\subsection{Experiments}

To answer the proposed research questions, we compare four HornFuzz configurations. We can describe these configurations using the set of options with which the fuzzer was launched~(\ref{tab:configurations}).
\begin{table}[tb]
  \caption{HornFuzz configurations.}
  \label{tab:configurations}
  \begin{tabular}{c|c|c} 
    \toprule
    \diagbox[width=13em]{-mutations}{-heuristic} & default & rare transitions \\\midrule
    our own mutations,& \multirow{2}{*}{naive} & \multirow{2}{*}{transition} \\
    Z3 rewrites & & \\\midrule
    all mutations: our own & \multirow{3}{*}{parameter} & \multirow{3}{*}{full} \\
    mutations, Z3 rewrites, & &  \\
    changing solver parameters & & \\
    \bottomrule
  \end{tabular}
\end{table}

\begin{itemize}
    \item A ``naive'' HornFuzz configuration does not use solver parameters and does not prioritize instances.
    \item A ``parameter'' configuration uses all mutations, but does not prioritize instances.
    \item A ``transition'' configuration does not use solver parameters, but prioritizes instances by rare transitions.
    \item A ``full'' HornFuzz configuration works with all mutations and prioritizes instances by rare transitions.
\end{itemize}

All of these configurations used a weighted choice of mutations. We compare these HornFuzz versions on 10 runs, each of which lasted 24 hours.

\textbf{RQ 1.} The code coverage and unique trace statistics averaged on all launches along with standard deviations are presented in table~\eqref{tab:coverage}. As a baseline, we compare against run with no mutations at all (i.e. against the basic unmutated seeds).

The data shows the configuration with seed selection by rare transitions outperforms the naive one. The transition version has more coverage on average (relative to the baseline) and also discovers almost the same number of unique traces, while having 17\% fewer executions. This can be explained by the fact that the launch of instances with more rare transitions is targeted at opening new unique traces. The connection between the focus on rare transitions and the coverage growth is not so obvious, but it is there: rare transitions often lead to previously unvisited solver code lines.

The version using solver parameters also outperforms the naive version in fewer launches. On average, it requires 5.88 runs to find a new unique trace, compared to 11.9 runs for the naive configuration. Thus, although the naive version of HornFuzz is faster, the transition and parameter versions are better at analyzing the solver behavior. 

The full HornFuzz configuration explores the solver much more efficiently than the naive one and the transition configuration. Compared to the parameter version, the advantage of the full version is not so significant, but still there and with fewer runs. Although the coverage of parameter version may be larger (considering the deviation), the full version is more stable. Thus, we can conclude that the use of solver parameters and seed selection based on rare transitions increases the HornFuzz efficiency.

\textbf{RQ 2:} To answer this question, we collected statistics on mutation weights. The mutation weight means the probability of opening a new trace when using this mutation. 

Table \eqref{tab:params} shows the solver parameter mutations with the highest weights for HornFuzz versions that use this mutation group. Table \eqref{tab:other_mut} shows the other mutations with the highest weights for all HornFuzz configurations. For each run, the final mutation weights were taken and then averaged over all runs.

\begin{table}[tb]
\begin{center}
\caption{Probabilities of opening a new trace when changing solver parameters.}
\label{tab:params}
\begin{tabular}{l|c|c}
    \toprule
    Parameter & Parameter & Full \\
    & version & version \\
    \midrule
    xform.transform\_arrays & 0.967 & 0.964 \\\midrule
    xform.slice & 0.957 & 0.960\\\midrule
    xform.inline\_eager & 0.750 & 0.761 \\\midrule
    xform.inline\_linear & 0.455 & 0.477 \\\midrule
    xform.tail\_simplifier\_pve & 0.376 & 0.399 \\\midrule
    xform.elim\_term\_ite & 0.321 & 0.320 \\\midrule
    xform.inline\_linear\_branch & 0.284 & 0.318 \\\midrule
    spacer.eq\_prop & 0.284 & 0.305 \\\midrule
    spacer.use\_inductive\_generalizer & 0.279 & 0.318 \\\midrule
    other & \multicolumn{2}{c}{$< 0.3$}  \\
    \bottomrule
\end{tabular}
\end{center}
\end{table}

\begin{table}[tb]
\begin{center}
\caption{Probabilities of opening a new trace when applying simplifications and our mutations.}
\label{tab:other_mut}
\begin{tabular}{l|c|c|c|c} 
\toprule
    Mutation & Naive & Transition & Parameter & Full \\
    & version & version & version & version \\
    \midrule
    SWAP\_OR & 0.282 & 0.344 & 0.350 & 0.380 \\\midrule
    ADD\_INEQ & 0.226 & 0.270 & 0.262 & 0.279 \\\midrule
    SWAP\_AND & 0.207 & 0.247 & 0.240 & 0.255 \\\midrule
    DUP\_AND & 0.207 & 0.251 & 0.242 & 0.255 \\\midrule
    MIX\_BOUND\_ & 0.191 & 0.230 & 0.219 & 0.234 \\
    VARS & & & & \\\midrule
    BREAK\_AND & 0.120 & 0.143 & 0.125 & 0.140 \\\midrule
    empty\_simplify & 0.094 & 0.113 & 0.109 & 0.118\\\midrule
    elim\_and & 0.009 & 0.012 & 0.010 & 0.014 \\\midrule
    other & \multicolumn{4}{c}{$< 0.01$}  \\
\bottomrule
\end{tabular}
\end{center}
\end{table}

The data obtained confirm our hypothesis about the efficiency of the solver parameters. The high weights of the solver parameters mean that their use often leads to the discovery of a new trace. 

It is also shown that proposed CHC solver-specific mutations have a significantly higher probability of opening a new trace than any simplifications (SMT-specific equivalent rewrites). Among simplifications empty\_simplify has the highest new trace discover probability, that is, simplification with rewriting rules enabled by default, which basically perform Boolean formulae simplifications.

\textbf{RQ 3:} The considered runs revealed only cases of incorrect model generation. We are not yet able to classify all bugs but we have done some work on bug localization. We manually divided the bugs found into several groups according to the parts of the solver to which they belong. These groups are presented in table~\eqref{tab:bug_groups}. The table shows bug numbers averaged on all launches along with standard deviations.

When analyzing the detected bugs, we noticed that many of them are caused by transformations of CHC systems: in particular, by linear and eager rule inlinings. In addition to transformation groups, our attention was also attracted by bugs that affect the Spacer core. We also found bugs that belong to several parts of the solver and categorize them as "unclassified".
%
%

\begin{table*}[tb]
\begin{center}
  \caption{Bug classification.}
  \label{tab:bug_groups}
  \begin{tabular}{l|c|c|c|c}
  \toprule
    Group & Naive version & Transition version & Parameter version & Full version \\
    \midrule
    Linear rule inlining transformation & 0~$\pm$~0 & 0.17~$\pm$~0.41 & 29.7~$\pm$~10.59 & 48.5~$\pm$~16.66\\
    \midrule
    Eager rule inlining transformation  & 25~$\pm$~12.33 & 31.67~$\pm$~8.87 & 13.7~$\pm$~6.52 & 19.75~$\pm$~11.54\\
    \midrule
    Other transformations & 1.67~$\pm$~1.61 & 3.17~$\pm$~3.06 & 1.6~$\pm$~1.07 & 5.25~$\pm$~6.32\\
    \midrule
    Spacer core & 0~$\pm$~0 & 0~$\pm$~0 & 0.7~$\pm$~2.21 & 0.13~$\pm$~0.35\\
    \midrule
    Unclassified & 5.08~$\pm$~3.99 & 5.17~$\pm$~4.79 & 5.9~$\pm$~3.25 & 12.13~$\pm$~4.12\\
    \midrule \midrule
    Number of bugs discovered & 31.75~$\pm$~15.59 & 40.17~$\pm$~9.75 & 51.6~$\pm$~7.81 & 85.75~$\pm$~21.35\\
  \bottomrule
  \end{tabular}
\end{center}
\end{table*}

The full version is shown to outperform all other configurations in terms of the number of bugs discovered on average. Also, it is shown that the use of solver parameter mutation or seed prioritization increases the number of bugs found compared to the naive version.

Since many discovered bugs belong to linear and eager rule inlinings, it is interesting to compare HornFuzz configurations in terms of number of discovered bugs in other parts of the solver, as this shows how the fuzzer explores different solver parts. By detecting bugs in the Spacer core, the parameter version is in the lead; however, in other groups, the full fuzzer configuration outperforms all other configurations.

In summary, the data obtained convince us that the use of solver parameters and instance selection by rare transitions improve the HornFuzz quality.

\subsection{Discovered bugs}
During our experiments, HornFuzz has found 2 confirmed satisfiability bugs and 13 confirmed model generation problems in Spacer. 11 of them have already been fixed by the Z3 developers, while others are in the process of being fixed.

The following issues have been resolved.
\begin{itemize}
\item Duplication of the conjunction element (mutation DUP\_AND) changed the solving result from sat to unsat\footnote{\url{https://github.com/Z3Prover/z3/issues/5714} [accessed: \today]}.
\item When solving the instance with the fp.xform.array\_blast solver parameter, the solving result changed from sat to unsat\footnote{\url{https://github.com/Z3Prover/z3/issues/5833} [accessed: \today]}. When solving a system with the fp.xform.array\_blast, in instances in the theory of arrays pairs of equalities of the form $(v_1 = select\ A\ i_1) \wedge (v_2 = select\ A\ i_2)$ were replaced by $(i_1 = i_2 \rightarrow v_1 = v_2)$ (Ackermann reduction~\cite{Arrays}), where, $A$~--- array parameterized by the set of indices $I$ and the set of values $V$, $i_1, i_2 \in I$, $v_1, v_2 \in V$.
\item 9 cases of incorrect model generation\footnote{\url{https://github.com/Z3Prover/z3/issues/5858} [accessed: \today] \\
\url{https://github.com/Z3Prover/z3/issues/5862} [accessed: \today]\\
\url{https://github.com/Z3Prover/z3/issues/5863} [accessed: \today]\\
\url{https://github.com/Z3Prover/z3/issues/5865} [accessed: \today]\\
\url{https://github.com/Z3Prover/z3/issues/5866} [accessed: \today]\\
\url{https://github.com/Z3Prover/z3/issues/5869} [accessed: \today]\\
\url{https://github.com/Z3Prover/z3/issues/5874} [accessed: \today]\\
\url{https://github.com/Z3Prover/z3/issues/5882} [accessed: \today]\\
\url{https://github.com/Z3Prover/z3/issues/5903} [accessed: \today]}.
\end{itemize}

A group of cases of incorrect model generation is associated with bugs in clause transformations. 4 such bugs have not been fixed\footnote{\url{https://github.com/Z3Prover/z3/issues/5920} [accessed: \today]} since fixing them and preventing other bugs of this type requires significant and thoughtful changes in the solver.

Since bug deduplication has not yet been implemented in the fuzzer, the bugs found were reported sequentially: HornFuzz always tested the solver version in which the last detected bug was fixed.

\section{Related Work}\label{sec:related}
The closest related work for HornFuzz is research on fuzzing SMT solvers. There are many efficient SMT solver fuzzers: STORM~\cite{STORM}, BanditFuzz~\cite{BanditFuzz}, FuzzSMT~\cite{FuzzSMT}, Falcon~\cite{Falcon}, OpFuzz~\cite{OpFuzz}, YinYang~\cite{YinYang}, etc. 

STORM is an open-source black-box mutational fuzzer for detecting critical bugs in SMT solvers. STORM uses fragments of existing SMT instances to generate new inputs. The fuzzer generates instances that are satisfiable by construction and thus it solves the oracle problem. 

BanditFuzz is a multiagent reinforcement learning performance SMT fuzzer. BanditFuzz generates tests according to the grammar given to it and mutates them while maintaining the structure. The fuzzer examines which grammatical structures lead to bug occurrences. BanditFuzz is open-source.

FuzzSMT is a grammar-based black-box SMT fuzzer. \mbox{FuzzSMT} randomly generates syntactically valid SMT formulae in array or bit-vector theory in order to detect critical defects.

Falcon is a grammar-based generative fuzzer based on exploring the functionalities used by the SMT solver (configuration space). It learns the correlations between the generated inputs (formula space) and the configuration space and proposes a feedback-driven mechanism.

OpFuzz is a type-aware mutational SMT fuzzer. OpFuzz leverages type-aware operator mutation to generate test inputs and validates the results of the SMT solvers by comparing the results of two or more solvers and reporting their inconsistencies.

YinYang is a mutational fuzzer for SMT solvers based on Semantic Fusion methodology. It fuse two existing equisatisfiable formulae into a new formula that combines the structures of its ancestors in a novel manner and preserves the satisfiability by construction.

Unfortunately, none of the SMT solver fuzzers is suitable for fuzzing CHC solvers, e.g., finding satisfiability bugs and model generation problems. STORM generates incorrect systems, that is, it does not preserve the CHC structure. This fuzzer creates the input instances from fragments of formulae from initial data. It generates a random assignment of all free variables in the formulae. Then, STORM randomly selects some parts of the original formulae and, in accordance with their values in the considered interpretation, composes instances from these subformulae. Thus, the probability that STORM will generate a Constrained Horn Clause is extremely small. 

BanditFuzz, FuzzSMT and Falcon can generate a correct Constrained Horn Clause systems, but it is difficult to create various Constrained Horn Clauses from scratch. The main problem is the satisfiability of rule clause bodies and query clause body. Firstly, at least one rule body for each uninterpreted predicate must be satisfiable, or we will end up with a trivial "false" interpretation. Secondly, if we want to generate satisfiable CHC system, we must guarantee that all rule bodies are unsatisfiable with query clause body. Therefore, we need to synthesize such~\cite{Synthesis} query body. Such a synthesis problem is comparable in complexity to the solution of the CHC system.

In addition, if the generation-based fuzzer cannot synthesize formulae that are known to be satisfiable, then it must check the results that the solver under test produces. Thus, another CHC solver is required to check the generated systems. Moreover, such a solver may also be needed in the case when the satisfiability is known in order to make sure that the formula generator is correct. However, at the moment the capabilities of the CHC solvers are very different, that is, not all systems can be checked. 

OpFuzz is also inefficient in testing CHC solvers, since it uses other solvers as oracles.

YinYang is able to generate correct CHC systems. However, since it obtains them by combining other CHC systems, its mutants grow very quickly, which leads to timeouts when solving. In addition, when obtaining instances in this way, the solver still considers each subsystem independently, since there are no clauses with predicates from both systems. Thus, the fuzzer does not explore new solutions.

\section{Conclusion}\label{sec:conclusion}
In this paper we present HornFuzz, the first fuzzer for testing the CHC solvers. HornFuzz is mutational and is based on metamorphic testing. It utilizes best practices from the state-of-the-art fuzzing research: has several seed selection heuristics and uses weighted selection for mutations. We also implemented a specialized reducer based on [hierarchical] delta debugging. 

HornFuzz has found bugs in the Spacer solver and its developers acknowledged them as genuine and (in some cases) serious problems. Some bugs have already been fixed, while others are in the process of being fixed.

While we were interested in validating Spacer as the most used CHC solver, testing other solvers may be future work. Now it requires the highlevel instrumentation: trace collection must be added to the solver under test. You cannot also use mutations that change Z3 parameters. But other fuzzer components can be used without modification.

Moreover, an important task for the future is to add bug deduplication. Now HornFuzz cannot determine whether bugs have a common cause or not. One correction can fix many bugs, and if we report every bug we find, we will create a lot of inconvenience for solver developers. Without bug deduplication we have to wait until the bug we reported is fixed to see if the others are reproducible and it slows down the bug reporting process.

Also, it would be interesting to extend the fuzzer with new mutations.

\bibliographystyle{ACM-Reference-Format}
\bibliography{hornfuzz}


\begin{thebibliography}{30}


\ifx \showCODEN    \undefined \def \showCODEN     #1{\unskip}     \fi
\ifx \showDOI      \undefined \def \showDOI       #1{#1}\fi
\ifx \showISBNx    \undefined \def \showISBNx     #1{\unskip}     \fi
\ifx \showISBNxiii \undefined \def \showISBNxiii  #1{\unskip}     \fi
\ifx \showISSN     \undefined \def \showISSN      #1{\unskip}     \fi
\ifx \showLCCN     \undefined \def \showLCCN      #1{\unskip}     \fi
\ifx \shownote     \undefined \def \shownote      #1{#1}          \fi
\ifx \showarticletitle \undefined \def \showarticletitle #1{#1}   \fi
\ifx \showURL      \undefined \def \showURL       {\relax}        \fi
\providecommand\bibfield[2]{#2}
\providecommand\bibinfo[2]{#2}
\providecommand\natexlab[1]{#1}
\providecommand\showeprint[2][]{arXiv:#2}

\bibitem[CHC(2022)]%
        {CHC-COMP-github}
 \bibinfo{year}{2022}\natexlab{}.
\newblock \bibinfo{booktitle}{\emph{The benchmarks selected at CHC-COMP 2021}}.
\newblock
\urldef\tempurl%
\url{https://github.com/chc-comp/chc-comp21-benchmarks}
\showURL{%
Retrieved May 6, 2022 from \tempurl}


\bibitem[SV-(2022)]%
        {SV-COMP-gitlab}
 \bibinfo{year}{2022}\natexlab{}.
\newblock \bibinfo{booktitle}{\emph{The benchmarks selected at last SV-COMP}}.
\newblock
\urldef\tempurl%
\url{https://gitlab.com/sosy-lab/benchmarking/sv-benchmarks/-/tree/main/clausess}
\showURL{%
Retrieved May 6, 2022 from \tempurl}


\bibitem[Alur et~al\mbox{.}(2013)]%
        {Synthesis}
\bibfield{author}{\bibinfo{person}{Rajeev Alur}, \bibinfo{person}{Rastislav
  Bodik}, \bibinfo{person}{Garvit Juniwal}, \bibinfo{person}{Milo M.~K.
  Martin}, \bibinfo{person}{Mukund Raghothaman}, \bibinfo{person}{Sanjit~A.
  Seshia}, \bibinfo{person}{Rishabh Singh}, \bibinfo{person}{Armando
  Solar-Lezama}, \bibinfo{person}{Emina Torlak}, {and}
  \bibinfo{person}{Abhishek Udupa}.} \bibinfo{year}{2013}\natexlab{}.
\newblock \showarticletitle{Syntax-guided synthesis}. In
  \bibinfo{booktitle}{\emph{2013 Formal Methods in Computer-Aided Design}}.
  \bibinfo{pages}{1--8}.
\newblock
\urldef\tempurl%
\url{https://doi.org/10.1109/FMCAD.2013.6679385}
\showDOI{\tempurl}


\bibitem[Bj{\o}rner et~al\mbox{.}(2015)]%
        {CHC}
\bibfield{author}{\bibinfo{person}{Nikolaj Bj{\o}rner}, \bibinfo{person}{Arie
  Gurfinkel}, \bibinfo{person}{Ken McMillan}, {and} \bibinfo{person}{Andrey
  Rybalchenko}.} \bibinfo{year}{2015}\natexlab{}.
\newblock \bibinfo{booktitle}{\emph{Horn Clause Solvers for Program
  Verification}}.
\newblock \bibinfo{publisher}{Springer International Publishing},
  \bibinfo{address}{Cham}, \bibinfo{pages}{24--51}.
\newblock
\showISBNx{978-3-319-23534-9}
\urldef\tempurl%
\url{https://doi.org/10.1007/978-3-319-23534-9_2}
\showDOI{\tempurl}


\bibitem[Brummayer and Biere(2009)]%
        {FuzzSMT}
\bibfield{author}{\bibinfo{person}{Robert Brummayer} {and}
  \bibinfo{person}{Armin Biere}.} \bibinfo{year}{2009}\natexlab{}.
\newblock \showarticletitle{Fuzzing and delta-debugging SMT solvers}.
\newblock \bibinfo{journal}{\emph{ACM International Conference Proceeding
  Series}} (\bibinfo{date}{01} \bibinfo{year}{2009}), \bibinfo{pages}{1--5}.
\newblock
\showISBNx{978-1-60558-484-3}
\urldef\tempurl%
\url{https://doi.org/10.1145/1670412.1670413}
\showDOI{\tempurl}


\bibitem[Böhme et~al\mbox{.}(2019)]%
        {Traces}
\bibfield{author}{\bibinfo{person}{Marcel Böhme}, \bibinfo{person}{Van-Thuan
  Pham}, {and} \bibinfo{person}{Abhik Roychoudhury}.}
  \bibinfo{year}{2019}\natexlab{}.
\newblock \showarticletitle{Coverage-Based Greybox Fuzzing as Markov Chain}.
\newblock \bibinfo{journal}{\emph{IEEE Transactions on Software Engineering}}
  \bibinfo{volume}{45}, \bibinfo{number}{5} (\bibinfo{year}{2019}),
  \bibinfo{pages}{489--506}.
\newblock
\urldef\tempurl%
\url{https://doi.org/10.1109/TSE.2017.2785841}
\showDOI{\tempurl}


\bibitem[Champion et~al\mbox{.}(2018)]%
        {Benchmarks2}
\bibfield{author}{\bibinfo{person}{Adrien Champion}, \bibinfo{person}{Naoki
  Kobayashi}, {and} \bibinfo{person}{Ryosuke Sato}.}
  \bibinfo{year}{2018}\natexlab{}.
\newblock \showarticletitle{HoIce: An ICE-based non-linear Horn clause solver}.
  In \bibinfo{booktitle}{\emph{Asian Symposium on Programming Languages and
  Systems}}. Springer, \bibinfo{pages}{146--156}.
\newblock


\bibitem[Chen et~al\mbox{.}(2020)]%
        {Metamorphic}
\bibfield{author}{\bibinfo{person}{T.~Y. Chen}, \bibinfo{person}{S.~C. Cheung},
  {and} \bibinfo{person}{S.~M. Yiu}.} \bibinfo{year}{2020}\natexlab{}.
\newblock \bibinfo{title}{Metamorphic Testing: A New Approach for Generating
  Next Test Cases}.
\newblock
\newblock
\urldef\tempurl%
\url{https://doi.org/10.48550/ARXIV.2002.12543}
\showDOI{\tempurl}


\bibitem[de~Moura and Bj{\o}rner(2008)]%
        {Z3}
\bibfield{author}{\bibinfo{person}{Leonardo de Moura} {and}
  \bibinfo{person}{Nikolaj Bj{\o}rner}.} \bibinfo{year}{2008}\natexlab{}.
\newblock \showarticletitle{Z3: An Efficient SMT Solver}. In
  \bibinfo{booktitle}{\emph{Tools and Algorithms for the Construction and
  Analysis of Systems}}, \bibfield{editor}{\bibinfo{person}{C.~R. Ramakrishnan}
  {and} \bibinfo{person}{Jakob Rehof}} (Eds.). \bibinfo{publisher}{Springer
  Berlin Heidelberg}, \bibinfo{address}{Berlin, Heidelberg},
  \bibinfo{pages}{337--340}.
\newblock
\showISBNx{978-3-540-78800-3}


\bibitem[Donaldson et~al\mbox{.}(2021)]%
        {Reduction}
\bibfield{author}{\bibinfo{person}{Alastair~F. Donaldson},
  \bibinfo{person}{Paul Thomson}, \bibinfo{person}{Vasyl Teliman},
  \bibinfo{person}{Stefano Milizia}, \bibinfo{person}{Andr\'{e}~Perez Maselco},
  {and} \bibinfo{person}{Antoni Karpi\'{n}ski}.}
  \bibinfo{year}{2021}\natexlab{}.
\newblock \showarticletitle{Test-Case Reduction and Deduplication Almost for
  Free with Transformation-Based Compiler Testing}. In
  \bibinfo{booktitle}{\emph{Proceedings of the 42nd ACM SIGPLAN International
  Conference on Programming Language Design and Implementation}} (Virtual,
  Canada) \emph{(\bibinfo{series}{PLDI 2021})}. \bibinfo{publisher}{Association
  for Computing Machinery}, \bibinfo{address}{New York, NY, USA},
  \bibinfo{pages}{1017–1032}.
\newblock
\showISBNx{9781450383912}
\urldef\tempurl%
\url{https://doi.org/10.1145/3453483.3454092}
\showDOI{\tempurl}


\bibitem[Fedyukovich and Rümmer(2021)]%
        {CHC-COMP-21}
\bibfield{author}{\bibinfo{person}{Grigory Fedyukovich} {and}
  \bibinfo{person}{Philipp Rümmer}.} \bibinfo{year}{2021}\natexlab{}.
\newblock \showarticletitle{Competition Report: CHC-COMP-21}.
\newblock \bibinfo{journal}{\emph{Electronic Proceedings in Theoretical
  Computer Science}}  \bibinfo{volume}{344} (\bibinfo{date}{Sep}
  \bibinfo{year}{2021}), \bibinfo{pages}{91–108}.
\newblock
\showISSN{2075-2180}
\urldef\tempurl%
\url{https://doi.org/10.4204/eptcs.344.7}
\showDOI{\tempurl}


\bibitem[Gurfinkel and Bjørner(2019)]%
        {CHC-tutorial}
\bibfield{author}{\bibinfo{person}{Arie Gurfinkel} {and}
  \bibinfo{person}{Nikolaj Bjørner}.} \bibinfo{year}{2019}\natexlab{}.
\newblock \showarticletitle{The Science, Art, and Magic of Constrained Horn
  Clauses}. In \bibinfo{booktitle}{\emph{2019 21st International Symposium on
  Symbolic and Numeric Algorithms for Scientific Computing (SYNASC)}}.
  \bibinfo{pages}{6--10}.
\newblock
\urldef\tempurl%
\url{https://doi.org/10.1109/SYNASC49474.2019.00010}
\showDOI{\tempurl}


\bibitem[Hojjat and Rümmer(2018)]%
        {Eldarica}
\bibfield{author}{\bibinfo{person}{Hossein Hojjat} {and}
  \bibinfo{person}{Philipp Rümmer}.} \bibinfo{year}{2018}\natexlab{}.
\newblock \showarticletitle{The ELDARICA Horn Solver}. In
  \bibinfo{booktitle}{\emph{2018 Formal Methods in Computer Aided Design
  (FMCAD)}}. \bibinfo{pages}{1--7}.
\newblock
\urldef\tempurl%
\url{https://doi.org/10.23919/FMCAD.2018.8603013}
\showDOI{\tempurl}


\bibitem[Komuravelli et~al\mbox{.}(2015)]%
        {Arrays}
\bibfield{author}{\bibinfo{person}{Anvesh Komuravelli},
  \bibinfo{person}{Nikolaj Bjorner}, \bibinfo{person}{Arie Gurfinkel}, {and}
  \bibinfo{person}{Kenneth~L. McMillan}.} \bibinfo{year}{2015}\natexlab{}.
\newblock \bibinfo{title}{Compositional Verification of Procedural Programs
  using Horn Clauses over Integers and Arrays}.
\newblock
\newblock
\urldef\tempurl%
\url{https://doi.org/10.48550/ARXIV.1508.01288}
\showDOI{\tempurl}


\bibitem[Komuravelli et~al\mbox{.}(2014a)]%
        {Spacer}
\bibfield{author}{\bibinfo{person}{Anvesh Komuravelli}, \bibinfo{person}{Arie
  Gurfinkel}, {and} \bibinfo{person}{Sagar Chaki}.}
  \bibinfo{year}{2014}\natexlab{a}.
\newblock \bibinfo{title}{SMT-based Model Checking for Recursive Programs}.
\newblock
\newblock
\showeprint[arxiv]{1405.4028}~[cs.LO]


\bibitem[Komuravelli et~al\mbox{.}(2014b)]%
        {MBP}
\bibfield{author}{\bibinfo{person}{Anvesh Komuravelli}, \bibinfo{person}{Arie
  Gurfinkel}, {and} \bibinfo{person}{Sagar Chaki}.}
  \bibinfo{year}{2014}\natexlab{b}.
\newblock \showarticletitle{SMT-Based Model Checking for Recursive Programs}.
  In \bibinfo{booktitle}{\emph{Computer Aided Verification}},
  \bibfield{editor}{\bibinfo{person}{Armin Biere} {and}
  \bibinfo{person}{Roderick Bloem}} (Eds.). \bibinfo{publisher}{Springer
  International Publishing}, \bibinfo{address}{Cham}, \bibinfo{pages}{17--34}.
\newblock
\showISBNx{978-3-319-08867-9}


\bibitem[Mansur et~al\mbox{.}(2020)]%
        {STORM}
\bibfield{author}{\bibinfo{person}{Muhammad~Numair Mansur},
  \bibinfo{person}{Maria Christakis}, \bibinfo{person}{Valentin W\"{u}stholz},
  {and} \bibinfo{person}{Fuyuan Zhang}.} \bibinfo{year}{2020}\natexlab{}.
\newblock \showarticletitle{Detecting Critical Bugs in SMT Solvers Using
  Blackbox Mutational Fuzzing}. In \bibinfo{booktitle}{\emph{Proceedings of the
  28th ACM Joint Meeting on European Software Engineering Conference and
  Symposium on the Foundations of Software Engineering}} (Virtual Event, USA)
  \emph{(\bibinfo{series}{ESEC/FSE 2020})}. \bibinfo{publisher}{Association for
  Computing Machinery}, \bibinfo{address}{New York, NY, USA},
  \bibinfo{pages}{701–712}.
\newblock
\showISBNx{9781450370431}
\urldef\tempurl%
\url{https://doi.org/10.1145/3368089.3409763}
\showDOI{\tempurl}


\bibitem[Misherghi and Su(2006)]%
        {HDD}
\bibfield{author}{\bibinfo{person}{Ghassan Misherghi} {and}
  \bibinfo{person}{Zhendong Su}.} \bibinfo{year}{2006}\natexlab{}.
\newblock \showarticletitle{HDD: Hierarchical Delta Debugging}. In
  \bibinfo{booktitle}{\emph{Proceedings of the 28th International Conference on
  Software Engineering}} (Shanghai, China) \emph{(\bibinfo{series}{ICSE '06})}.
  \bibinfo{publisher}{Association for Computing Machinery},
  \bibinfo{address}{New York, NY, USA}, \bibinfo{pages}{142–151}.
\newblock
\showISBNx{1595933751}
\urldef\tempurl%
\url{https://doi.org/10.1145/1134285.1134307}
\showDOI{\tempurl}


\bibitem[Mordvinov and Fedyukovich(2017)]%
        {Benchmarks1}
\bibfield{author}{\bibinfo{person}{Dmitry Mordvinov} {and}
  \bibinfo{person}{Grigory Fedyukovich}.} \bibinfo{year}{2017}\natexlab{}.
\newblock \bibinfo{title}{Verifying Safety of Functional Programs with
  Rosette/Unbound}.
\newblock
\newblock
\urldef\tempurl%
\url{https://doi.org/10.48550/ARXIV.1704.04558}
\showDOI{\tempurl}


\bibitem[Mordvinov and Fedyukovich(2019)]%
        {Dima}
\bibfield{author}{\bibinfo{person}{Dmitry Mordvinov} {and}
  \bibinfo{person}{Grigory Fedyukovich}.} \bibinfo{year}{2019}\natexlab{}.
\newblock \showarticletitle{Property Directed Inference of Relational
  Invariants}. In \bibinfo{booktitle}{\emph{2019 Formal Methods in Computer
  Aided Design (FMCAD)}}. \bibinfo{pages}{152--160}.
\newblock
\urldef\tempurl%
\url{https://doi.org/10.23919/FMCAD.2019.8894274}
\showDOI{\tempurl}


\bibitem[Norris(1997)]%
        {Markov_chain}
\bibfield{author}{\bibinfo{person}{J.~R. Norris}.}
  \bibinfo{year}{1997}\natexlab{}.
\newblock \bibinfo{booktitle}{\emph{Markov Chains}}.
\newblock \bibinfo{publisher}{Cambridge University Press}.
\newblock
\urldef\tempurl%
\url{https://doi.org/10.1017/CBO9780511810633}
\showDOI{\tempurl}


\bibitem[Offutt and Xu(2004)]%
        {Mutation-based}
\bibfield{author}{\bibinfo{person}{Jeff Offutt} {and} \bibinfo{person}{Wuzhi
  Xu}.} \bibinfo{year}{2004}\natexlab{}.
\newblock \showarticletitle{Generating Test Cases for Web Services Using Data
  Perturbation}.
\newblock \bibinfo{journal}{\emph{SIGSOFT Softw. Eng. Notes}}
  \bibinfo{volume}{29}, \bibinfo{number}{5} (\bibinfo{date}{sep}
  \bibinfo{year}{2004}), \bibinfo{pages}{1–10}.
\newblock
\showISSN{0163-5948}
\urldef\tempurl%
\url{https://doi.org/10.1145/1022494.1022529}
\showDOI{\tempurl}


\bibitem[Pacheco et~al\mbox{.}(2007)]%
        {Coverage}
\bibfield{author}{\bibinfo{person}{Carlos Pacheco},
  \bibinfo{person}{Shuvendu~K. Lahiri}, \bibinfo{person}{Michael~D. Ernst},
  {and} \bibinfo{person}{Thomas Ball}.} \bibinfo{year}{2007}\natexlab{}.
\newblock \showarticletitle{Feedback-Directed Random Test Generation}. In
  \bibinfo{booktitle}{\emph{29th International Conference on Software
  Engineering (ICSE'07)}}. \bibinfo{pages}{75--84}.
\newblock
\urldef\tempurl%
\url{https://doi.org/10.1109/ICSE.2007.37}
\showDOI{\tempurl}


\bibitem[Satake et~al\mbox{.}(2020)]%
        {PCSat}
\bibfield{author}{\bibinfo{person}{Yuki Satake}, \bibinfo{person}{Hiroshi
  Unno}, {and} \bibinfo{person}{Hinata Yanagi}.}
  \bibinfo{year}{2020}\natexlab{}.
\newblock \showarticletitle{Probabilistic Inference for Predicate Constraint
  Satisfaction}.
\newblock \bibinfo{journal}{\emph{Proceedings of the AAAI Conference on
  Artificial Intelligence}}  \bibinfo{volume}{34} (\bibinfo{date}{04}
  \bibinfo{year}{2020}), \bibinfo{pages}{1644--1651}.
\newblock
\urldef\tempurl%
\url{https://doi.org/10.1609/aaai.v34i02.5526}
\showDOI{\tempurl}


\bibitem[Scott et~al\mbox{.}(2021)]%
        {BanditFuzz}
\bibfield{author}{\bibinfo{person}{Joseph Scott}, \bibinfo{person}{Trishal
  Sudula}, \bibinfo{person}{Hammad Rehman}, \bibinfo{person}{Federico Mora},
  {and} \bibinfo{person}{Vijay Ganesh}.} \bibinfo{year}{2021}\natexlab{}.
\newblock \showarticletitle{BanditFuzz: Fuzzing SMT Solvers with Multi-agent
  Reinforcement Learning}. In \bibinfo{booktitle}{\emph{Formal Methods}},
  \bibfield{editor}{\bibinfo{person}{Marieke Huisman}, \bibinfo{person}{Corina
  P{\u{a}}s{\u{a}}reanu}, {and} \bibinfo{person}{Naijun Zhan}} (Eds.).
  \bibinfo{publisher}{Springer International Publishing},
  \bibinfo{address}{Cham}, \bibinfo{pages}{103--121}.
\newblock
\showISBNx{978-3-030-90870-6}


\bibitem[Sutton et~al\mbox{.}(2007)]%
        {Fuzzing}
\bibfield{author}{\bibinfo{person}{Michael Sutton}, \bibinfo{person}{Adam
  Greene}, {and} \bibinfo{person}{Pedram Amini}.}
  \bibinfo{year}{2007}\natexlab{}.
\newblock \bibinfo{booktitle}{\emph{Fuzzing: brute force vulnerability
  discovery}}.
\newblock \bibinfo{publisher}{Pearson Education}.
\newblock


\bibitem[Winterer and Zhang(2020)]%
        {YinYang}
\bibfield{author}{\bibinfo{person}{Dominik Winterer} {and}
  \bibinfo{person}{Chengyu Zhang}.} \bibinfo{year}{2020}\natexlab{}.
\newblock \showarticletitle{Validating SMT solvers via semantic fusion}. In
  \bibinfo{booktitle}{\emph{Proceedings of the 41st ACM SIGPLAN Conference on
  Programming Language Design and Implementation}}. \bibinfo{pages}{718--730}.
\newblock


\bibitem[Winterer et~al\mbox{.}(2020)]%
        {OpFuzz}
\bibfield{author}{\bibinfo{person}{Dominik Winterer}, \bibinfo{person}{Chengyu
  Zhang}, {and} \bibinfo{person}{Zhendong Su}.}
  \bibinfo{year}{2020}\natexlab{}.
\newblock \showarticletitle{On the Unusual Effectiveness of Type-Aware Operator
  Mutations for Testing SMT Solvers}.
\newblock \bibinfo{journal}{\emph{Proc. ACM Program. Lang.}}
  \bibinfo{volume}{4}, \bibinfo{number}{OOPSLA}, Article
  \bibinfo{articleno}{193} (\bibinfo{date}{nov} \bibinfo{year}{2020}),
  \bibinfo{numpages}{25}~pages.
\newblock
\urldef\tempurl%
\url{https://doi.org/10.1145/3428261}
\showDOI{\tempurl}


\bibitem[Yao et~al\mbox{.}(2021)]%
        {Falcon}
\bibfield{author}{\bibinfo{person}{Peisen Yao}, \bibinfo{person}{Heqing Huang},
  \bibinfo{person}{Wensheng Tang}, \bibinfo{person}{Qingkai Shi},
  \bibinfo{person}{Rongxin Wu}, {and} \bibinfo{person}{Charles Zhang}.}
  \bibinfo{year}{2021}\natexlab{}.
\newblock \showarticletitle{Fuzzing SMT Solvers via Two-Dimensional Input Space
  Exploration}. In \bibinfo{booktitle}{\emph{Proceedings of the 30th ACM
  SIGSOFT International Symposium on Software Testing and Analysis}} (Virtual,
  Denmark) \emph{(\bibinfo{series}{ISSTA 2021})}.
  \bibinfo{publisher}{Association for Computing Machinery},
  \bibinfo{address}{New York, NY, USA}, \bibinfo{pages}{322–335}.
\newblock
\showISBNx{9781450384599}
\urldef\tempurl%
\url{https://doi.org/10.1145/3460319.3464803}
\showDOI{\tempurl}


\bibitem[Yoo and Harman(2012)]%
        {Selection}
\bibfield{author}{\bibinfo{person}{S. Yoo} {and} \bibinfo{person}{M. Harman}.}
  \bibinfo{year}{2012}\natexlab{}.
\newblock \showarticletitle{Regression Testing Minimization, Selection and
  Prioritization: A Survey}.
\newblock \bibinfo{journal}{\emph{Softw. Test. Verif. Reliab.}}
  \bibinfo{volume}{22}, \bibinfo{number}{2} (\bibinfo{date}{mar}
  \bibinfo{year}{2012}), \bibinfo{pages}{67–120}.
\newblock
\showISSN{0960-0833}
\urldef\tempurl%
\url{https://doi.org/10.1002/stv.430}
\showDOI{\tempurl}


\end{thebibliography}

\appendix
\section{Mutations}
\subsection{Simplify parameters}\label{app:simplify}
\begin{enumerate}
	\item arith\_ineq\_lhs;
	\item arith\_lhs; 
	\item blast\_distinct; 
	\item blast\_select\_store; 
	\item elim\_and;
	\item elim\_rem; 
	\item elim\_to\_real;
	\item eq2ineq; 
	\item expand\_power; 
	\item expand\_select\_ite; 
	\item expand\_select\_store; 
	\item expand\_store\_eq; 
	\item expand\_tan; 
	\item gcd\_rounding; 
	\item hoist\_ite; 
	\item hoist\_mul;
	\item ite\_extra\_rules;
	\item local\_ctx;
	\item mul2concat;
	\item mul\_to\_power;
	\item pull\_cheap\_ite;
	\item push\_ite\_arith; 
	\item rewrite\_patterns; 
	\item som; 
	\item sort\_store; 
	\item sort\_sums;
	\item split\_concat\_eq; 
	\item algebraic\_number\_evaluator; 
	\item elim\_ite;
	\item elim\_sign\_ext;
	\item flat;
	\item push\_to\_real;
 		\item ignore\_patterns\_on\_ground\_qbody.
 \end{enumerate}
\subsection{Solver parameters}\label{app:parameters}
\begin{enumerate}
	\item spacer.ctp;
    \item spacer.elim\_aux;
    \item spacer.eq\_prop;
    \item spacer.ground\_pobs;
    \item spacer.keep\_proxy;
    \item spacer.mbqi;
    \item spacer.propagate;
    \newpage
    \item spacer.reach\_dnf;
    \item spacer.use\_array\_eq\_generalizer;
    \item spacer.use\_derivations;
    \item spacer.use\_inc\_clause;
    \item spacer.use\_inductive\_generalizer;
    \item xform.coi;
    \item xform.compress\_unbound;
    \item xform.inline\_eager;
    \item xform.inline\_linear;
    \item xform.slice;
    \item xform.tail\_simplifier\_pve.
    \item spacer.p3.share\_invariants;
    \item spacer.p3.share\_lemmas;
    \item spacer.use\_lim\_num\_gen;
    \item spacer.reset\_pob\_queue;
    \item spacer.simplify\_lemmas\_post;
    \item spacer.simplify\_lemmas\_pre;
    \item spacer.simplify\_pob;
    \item spacer.use\_bg\_invs;
    \item spacer.use\_euf\_gen;
    \item spacer.use\_lemma\_as\_cti;
    \item xform.array\_blast\_full;
    \item xform.coalesce\_rules;
    \item xform.elim\_term\_ite;
    \item xform.inline\_linear\_branch;
    \item xform.instantiate\_arrays;
    \item xform.instantiate\_arrays.enforce;
    \item xform.instantiate\_quantifiers;
    \item xform.quantify\_arrays;
    \item xform.transform\_arrays.
 \end{enumerate}

\end{document}